\documentclass[english,amsmath,showpacs,amssymb,aps,twocolumn,pra]{revtex4}
\usepackage[T1]{fontenc}
\usepackage[utf8]{inputenc}
\usepackage{graphicx}
\usepackage{amscd}
\usepackage{mathrsfs}
\usepackage{amsfonts}
\usepackage{dcolumn}
\usepackage{color}
\usepackage{bm}

\usepackage{ulem}
\newcommand\redout{\bgroup\markoverwith
{\textcolor{red}{\rule[.5ex]{2pt}{1.5pt}}}\ULon}
\usepackage{soul}
\setstcolor{red}
\setulcolor{red}

\begin{document}
\global\long\def\ra{\rangle}
 \global\long\def\la{\langle}
 \global\long\def\pp{\partial}
 \global\long\def\ov#1{\overline{#1}}
 \global\long\def\ul#1{\underline{#1}}

\global\long\def\ww#1{\widetilde{#1}}
 \global\long\def\bm{{\bf m}}
 \global\long\def\bn{{\bf n}}
 \global\long\def\bp{{\bf p}}

\global\long\def\bP{{\bf P}}
 \global\long\def\br{{\bf r}}
 \global\long\def\bR{{\bf R}}
 \global\long\def\bx{{\bf x}}
 \global\long\def\by{{\bf y}}

\global\long\def\E{{\cal E}}
 \global\long\def\lm{{\lambda}}
 \global\long\def\ve{\varepsilon}
 \global\long\def\mP{{\mathscr{P}}}
 \global\long\def\oa{\overline{\alpha}}

\global\long\def\C{{{\cal C}_{\lambda}}}
 \global\long\def\ora#1{\overrightarrow{#1}}
 \global\long\def\HH{\mathscr{H}}
 \global\long\def\tr{{\rm Tr}}

\global\long\def\var{{\rm var}\,}
 \global\long\def\im{{\rm i}}
 \global\long\def\mo{\mathcal{O}}
 \global\long\def\eq{\equiv}

\global\long\def\mt{\mathrm{T}}
\global\long\def\mh{\mathcal{H}}
\global\long\def\lr#1{\left[#1\right]}

\global\long\def\pb#1#2#3#4#5#6#7#8{\{\frac{\partial#5}{\partial#1}\frac{\partial#6}{\partial#2}-\frac{\partial#7}{\partial#3}\frac{\partial#8}{\partial#4}\}_{PB}}

\global\long\def\wq{\infty}
 \global\long\def\sde{\sum_{n=1}^{d_{\E}}}
 \global\long\def\fde{\frac{1}{d_{\E}}}
 \global\long\def\fds{\frac{1}{d_{S}}}

\global\long\def\hs{{H}_{S}}
 \global\long\def\he{{H}_{\E}}
 \global\long\def\his{{H}_{S}^{I}}
 \global\long\def\hie{{H}_{\E}^{I}}
 \global\long\def\hh#1#2{ {H}_{#1}^{#2}}

\global\long\def\hsw{{ H}_{S}}
 \global\long\def\hiew{{ H}_{\E}^{I}}
\global\long\def\hhwn#1#2{ h_{#1}^{#2}}
 \global\long\def\si{\sigma_{i}}
 \global\long\def\ssi{\sigma_{i}^{2}}

\global\long\def\bsi{|S_{i}\ra}
 \global\long\def\bsip{|S_{i}^{\prime}\ra}
 \global\long\def\bs#1{|S_{#1}\ra}
 \global\long\def\sums#1#2{\sum_{#1}^{#2}}

\global\long\def\de{d_{\E}}
 \global\long\def\ds{d_{S}}
 \global\long\def\stde#1{\sum_{#1 =1}^{\de}}
 \global\long\def\mbx#1{\mbox{#1}}

 \global\long\def\tcb#1{\textcolor{blue}{#1}}

\newcommand{\be}{\begin{equation}}
\newcommand{\ee}{\end{equation}}
\newcommand{\bey}{\begin{eqnarray}}
\newcommand{\eey}{\end{eqnarray}}
\newcommand{\nn}{\nonumber}
\global\long\def\vec#1{\mbox{\boldsymbol{$#1$}}}
\newcommand{\ssst}{\scriptscriptstyle}
\newcommand{\sst}{\scriptstyle}

 \title {Statistically preferred basis of an open quantum system: Its
  relation to the eigenbasis of a renormalized self-Hamiltonian
}

 \author{Lewei He and Wen-ge Wang\footnote{ Email address: wgwang@ustc.edu.cn}}

 \affiliation{
 Department of Modern Physics, University of Science and Technology of China,
 Hefei 230026, China
 }

 \date{\today}

 \begin{abstract}

 We study the problem of the basis of an open quantum system, under a quantum chaotic environment, which is preferred in view of its stationary reduced density matrix (RDM), that is, the basis in which the stationary RDM is diagonal. It is shown that, under an initial condition composed of sufficiently many energy eigenstates of the total system, such a basis is given by the eigenbasis of a renormalized self-Hamiltonian of the system, in the limit of large Hilbert space of the environment. Here, the renormalized self-Hamiltonian is given by the unperturbed self-Hamiltonian plus a certain average of the interaction Hamiltonian over the environmental degrees of freedom.
 Numerical simulations, performed in two models, both with the kicked rotor as the environment, give results consistent with the above analytical predictions for the limit of large environment.

 \end{abstract}



 \maketitle

\section{Introduction}

 The reduced density matrix (RDM) is of central importance
 in understanding properties of open quantum systems.
 An important topic is the condition under which Schr\"{o}dinger evolution of a total
 system may bring the RDM of a subsystem into a stationary solution.
 In the case that a stationary RDM exists, a further important topic is
 the basis in which the stationary RDM becomes diagonal, as well as properties of its diagonal
 elements in such a basis.

 Recently, concerning the above problems, impressive progresses have been seen in the
 field of the foundation of statistical physics
 {\cite{speed,PSW06,LPSW09,Goldstein06,Eisert12,pre-sms-12,Reimann,Short,Gogolin10,LCG12,MPCP13,Wu13} }.
 It has been shown that typical vectors within a certain energy shell of a total system
 give almost the same RDM for a given subsystem,
 provided that the dimension of the Hilbert space of the subsystem is sufficiently small
 compared with that of the energy shell \cite{PSW06}.
 Since a typical vector may typically evolve  into another typical vector,
 this result implies that the RDM may become almost stationary,
 once the total system has reached a typical state.
 The condition for the appearance of an almost stationary RDM can be even further relaxed
 {\cite{LPSW09, Short, Reimann}}.
 Moreover, under weak system-environment interaction,
 the RDM is almost diagonal in the eigenbasis of the self-Hamiltonian of the subsystem,
 with diagonal elements having the canonical distribution \cite{Goldstein06,Eisert12}.
 Further, similar results can also be obtained for relatively weak system-environment interaction,
 with a renormalized self-Hamiltonian that appropriately takes into account
 the impact of the system-environment interaction \cite{pre-sms-12}.

 The problem of the existence of a somewhat fixed basis in which the RDM may become diagonal is
 also of interest in the field of decoherence, where
 it is called a preferred pointer basis
 \cite{Zeh70,DK00,Zurek03,Schloss04,JZKGKS03,Zurek81,PZ99,WGCL08,BHS01,WHG12},
 a concept which was originally introduced to capture the
 robustness of certain properties of macroscopic objects
 like pointers in measurement instruments \cite{Zeh70,Zurek81}
 and is nowadays also used at the microscopic level.
 Under weak system-environment coupling, the eigenstates of
 the system's self-Hamiltonian may form a good preferred basis \cite{PZ99,WGCL08}.
 Under strong system-environment interaction,
 when the influence of the self-Hamiltonian can be neglected, e.g., for short times,
 the eigenstates of the interaction Hamiltonian are ``preferred'' \cite{Zurek81,BHS01};
 meanwhile, observed numerically, such eigenstates may be
 preferred even for long times in certain models \cite{WHG12}.
 The case of intermediate interaction strength is more complex.
 Still, it has been observed numerically that preferred states may exist, changing
 continuously from eigenstates of the self-Hamiltonian
 to those of the interaction Hamiltonian with increasing coupling strength \cite{WHG12}.

 In this paper, we are to investigate whether or not a uniform picture
 may be available for the basis that is preferred in view of the long-time evolution of the RDM,
 in the whole regime of the system-environment interaction strength.
 Since the stationariness of RDM does not necessarily imply all the features that are usually
 expected for a preferred pointer basis,
 we use the term {\it statistically preferred basis} (SPB)
 to refer to a fixed basis, in which the RDM may approach a diagonal form when it becomes stationary.
 At first sight, there seems no simple answer to this question, when the system's self-Hamiltonian
 is not commutable with the interaction Hamiltonian.
 However, in this paper, we show that a simple solution may indeed exist,
 if the environment is sufficiently large and undergoes a sufficiently chaotic
 motion to be specified below.
 Indeed, as already known, certain chaotic or random properties of the environment may lead to
 equilibration or thermalization of open quantum systems
 \cite{Peres84,Deutsch,Srednicki94,Rigol,Ikeda, Altland-Haake,Marko,Brandao,Masanes,Ududec,VZ12}.

 The paper is structured as follows.
 In Sec.\ref{sect-SPB}, we give the main analytical results, that is, when certain conditions are
 satisfied, a stationary RDM is diagonal in the eigenbasis of a renormalized
 self-Hamiltonian of the system.
 This implies that, if a SPB exists under the conditions,
 it is given by the eigenbasis of the renormalized self-Hamiltonian, regardless of the interaction strength.
 Section \ref{sect-numeric} is devoted to discussions of numerical simulations performed in two models,
 showing results consistent with the analytical predictions.
 Finally, conclusions and a brief discussion are given in Sec.\ref{sect-conclusion}.

\section{Statistically-preferred basis under a chaotic environment}
\label{sect-SPB}

 In this section, we give the analytical results of this paper.
 The settings, notations, and some preliminary discussions are given in Sec.~\ref{sect-notation}.
 The main result is given in Sec.~\ref{sect-theorem}, for a traceless interaction Hamiltonian
 with a product form.
 Then, in Sec.~\ref{sect-generic-HI}, it is shown that, for a generic interaction Hamiltonian,
 the above-mentioned result is still valid after a renormalization of the self- and interaction Hamiltonians.

 \subsection{Settings, notations, and preliminary arguments}
 \label{sect-notation}

 We consider a total system, which is composed of a central system $S$ and
 its environment denoted by $\E$.
 We use $\HH$, $\HH_S$, and $\HH_\E$ to denote the Hilbert spaces of the total system, the subsystem
 $S$, and the environment $\E$, respectively, and use $d_S$ and $d_\E$ to denote the dimensions of
 $\HH_S$ and $\HH_\E$, respectively.
 We assume that $d_S$ is negligibly small compared with $d_\E$;
 specifically, $d_S$ is always kept finite, while $d_\E$ may go to infinity.

 We assume that the environment $\E$ is a quantum chaotic system.
 The total Hamiltonian is written as
\begin{equation}\label{H}
 H = H_S + H_\E + H_I,
\end{equation}
 where $H_S$ and $H_\E$ represent the Hamiltonians of $S$ and $\E$, respectively, and $ H_I$
 indicates the interaction Hamiltonian.
 We consider a product form of $H_I$, namely,
\begin{equation}\label{HI}
 H_I = H^{IS} \otimes H^{I\E},
\end{equation}
 where $H^{IS}$ and $H^{I\E}$ are Hermitian operators acting on the two Hilbert spaces
 $\HH_S$ and $\HH_\E$, respectively.
 (A more generic form of $H_I$ will be discussed in Sec.~\ref{sect-generic-HI}.)
 Eigenstates of $H^{I\E}$ will be used in our discussions and will be denoted by $|\varphi_n\ra $
 with eigenvalues $h_{n}^{I}$, which may have some degeneracy, i.e.,
\begin{equation}\label{theta-n}
 {H}^{I\E}|\varphi_{n}\ra= h_n^I |\varphi_{n}\ra.
\end{equation}
 We assume that the operator $H^{I\E}$ is bounded and
 has a discrete spectrum, even in the limit of large environment.

 The evolution of a generic, normalized state vector $|\psi (t)\ra $ in the total Hilbert space
 obeys the Schr\"{o}dinger equation,
\begin{equation}\label{SE}
 i \hbar \frac{d}{dt}|\psi (t)\ra = H |\psi(t)\ra .
\end{equation}
 The RDM of the system $S$, denoted by $\rho^S \equiv \tr_{\E} \rho$ with
 $\rho (t) = |\psi(t)\ra \la \psi(t)|$, satisfies the following equation:
\begin{equation}\label{drhos-dt}
 i \hbar \frac{d{\rho}^{S}(t)}{dt}= - \tr_{\E}(\left[ {\rho}(t),{H}\right]).
\end{equation}
 Below, we give intuitive arguments for a relation between the eigenstates of a
 stationary RDM  and the eigenstates of $H_S$.
 More rigorous discussions will be given in the next two subsections.

 Let us use $\{ |S_i\ra \}$ to denote a fixed, orthonormal basis in $\HH_S$.
 The state vector $|\psi(t)\ra$ is decomposed in the following way,
\begin{equation}\label{psi-expan}
 |\psi(t)\ra=\sum_{i=1}^{d_S}|S_{i}\ra|\E_{i}(t)\ra,
\end{equation}
 where $|\E_i(t)\ra \in \HH_\E$ are usually not normalized.
 The elements of $\rho^S$, defined by $\rho^S_{ij}(t) = \la S_i |\rho^S(t) |S_j\ra $,
 can be expressed in terms of the components $|\E_i(t)\ra $, namely,
\begin{equation}\label{rhosij-Eij}
    \rho^S_{ij}(t) = \la \E_j(t) |\E_i(t)\ra .
\end{equation}
 After some derivation (see Appendix~\ref{appendix-rhos} for details),
 it can be shown that the elements $\rho^S_{ij}(t)$ satisfy the following equation,
\begin{equation}
 -i \hbar \frac{d\rho_{ij}^{S}}{dt} = W^{(1)}_{ij} + W^{(2)}_{ij},
 \label{eq:change of rho(compact form)}
\end{equation}
 where
\begin{eqnarray}\label{W1}
 W^{(1)}_{ij} & = & \la S_i |\left[\rho^{S},H_{S}\right] |S_j \ra ,
 \\ W^{(2)}_{ij} & = & \sum_{q=1}^{d_{S}}H_{qj}^{IS}H_{qi}^{I\E}-\sum_{p=1}^{d_{S}}H_{ip}^{IS}H_{jp}^{I\E}. \label{W2}
\end{eqnarray}
 Here,
\begin{eqnarray} \label{HIS} H_{ij}^{IS} & \eq & \la S_{i}|{H}^{IS}|S_{j}\ra
 \\ H_{ij}^{I\E} & \eq & \la\E_{i}(t)|{H}^{I\E}|\E_{j}(t)\ra . \label{HIE}
\end{eqnarray}

 Suppose that the RDM may approach a stationary form, which is diagonal in the basis $\{ |S_i\ra\} $,
 i.e., $\rho^S(t)|S_i\ra  \simeq \rho^S_{ii} |S_i\ra $ for long times $t$.
 Then, for long times, Eq.(\ref{W1}) gives
\begin{equation}\label{W1-sim}
W^{(1)}_{ij} \simeq \la S_i|H_S|S_j\ra ( \rho^S_{ii}(t) - \rho^S_{jj}(t) ).
\end{equation}
 Note that (i) the stationariness of the RDM $\rho^S(t)$ {usually} requires vanishing $W^{(1)}_{ij}$ and 
 (ii) the diagonal elements of the RDM are usually initial-state dependent, hence are not necessarily
 the same.
 Then, Eq.~(\ref{W1-sim}) suggests that $H_S$ may have a diagonal form in the basis ${ \{|S_i\ra\} } $.

 \subsection{The RDM under a traceless interaction Hamiltonian}
 \label{sect-theorem}

 In this subsection, we discuss the central result of this paper.
 As is known, under an initial condition that involves many (quasi)energy
 eigenstates, beyond a certain time period, a chaotic system behaves quite irregularly and
 its wave function can be regarded as possessing certain random features
 \cite{Berry,CC94book}.
 In particular, when the random matrix theory can be assumed to be applicable,
 components of the wave functions can be regarded as Gaussian random variables
 \cite{Haake,pre02-mixed}.
 Below, exploiting this type of random feature, we study properties of the RDM of the central system.

 Let us expand the components $|\E_i(t)\ra $ in Eq.~(\ref{psi-expan})
 in the eigenbasis $|\varphi_{n}\ra$ of the operator $H^{I\E}$,
\begin{equation}\label{Ei-expan}
 |\E_{i}(t)\ra=\frac{1}{\sqrt{d_{\E}}}\sum_{n=1}^{d_{\E}}C_{in}(t)|\varphi_{n}\ra.
\end{equation}
 Initially, there may exist correlation among the coefficients $C_{in}(t)$.
 Due to the chaotic motion of the environment, the correlation among the coefficients decays fast.
 Beyond some time period, the correlation can be neglected
 and it would be reasonable to expected that the coefficients $C_{in}(t)$
 may be treated as random variables effectively.

 Now, we state the central analytical result of this paper.
 That is, if there exists a time scale $\tau_R$, such that
 the coefficients $C_{in}(t)$ for each time $t>\tau_R$ can be effectively treated as
 independent random variables possessing the property to be stated below, then,
\begin{equation} \label{eq:rho_hs_commute}
 \lim_{d_{\E}\to\infty} \left[ {H}_{S}, {\rho}^{S}(t)\right]=0 \quad \text{for} \ t > \tau_R.
\end{equation}
 Using $X_{in}$ to denote  the independent random variables mentioned above,
 the property is that they have mean zero, $\la X_{in}\ra=0$,
 time-independent and finite variances, $\la|X_{in}|^{2}\ra=\sigma_{i}^{2}$
 with $\sum_i \sigma_{i}^{2}=1$, and finite $\la|X_{in}|^{4}\ra$.
 Here and hereafter, we use $\la X \ra $ to indicate the statistical average of a
 random variable $X$.
 Since $\sigma_i$ are assumed to be independent of the time $t$,
 the time scale $\tau_R$ should be at least larger than the relaxation time
 [$\sigma_i^2$ giving diagonal elements of the stationary RDM as shown
 in Eq.(\ref{eq:equi-state_of_s}) given below].

 Below in this section, we show validity of Eq.(\ref{eq:rho_hs_commute}) in the case that
 the operator $H^{I\E}$ satisfies
\begin{equation} \label{hIE=0}
 \ov h^{I\E} \equiv \lim_{d_{\E}\to\infty} \tr_\E H^{I\E}/d_\E =0.
\end{equation}
 For this purpose, we make use of the following property of
 the random variables $X_{in}$ (see Appendix~\ref{app-t2} for the proof), that is,
\begin{equation} \label{HIE-CC}
 \lim_{d_{\E}\to\infty}  \frac{1}{d_{\E}}\sde h_n^I X_{jn}^{*}X_{in} =0 \quad \forall (i,j).
\end{equation}
 Since the coefficients $C_{in}(t)$ can be taken as the random variables $X_{in}$,
 noticing Eq.~(\ref{Ei-expan}), it is easy to verify that,
 as a consequence of Eq.~(\ref{HIE-CC}), $H^{I\E}_{ji}$ in Eq.~(\ref{HIE}) satisfies
\begin{equation}
 \lim_{d_{\E}\to\infty}  H^{I\E}_{ji}  = 0\
\quad \forall (i,j). \label{eq:hiew=00003D0}
\end{equation}
 Then, substituting Eq.~(\ref{Ei-expan}) into Eq.~(\ref{rhosij-Eij})
 and making use of the above-discussed properties of $X_{in}$, in particular, the time independency
 of $\sigma_i$, it is seen that the RDM is stationary for $t>\tau_R$ in the limit of large $d_{\E}$,
\begin{equation}\label{drhoS=0}
 \lim_{d_\E \to \infty} d\rho^{S}/dt=0 \quad \text{for} \ t > \tau_R,
\end{equation}
 with a diagonal form in the basis $\{|S_i\ra\} $,
\begin{eqnarray}
  \rho^{S}_{ij} =  \fde\sde C_{jn}^{*}C_{in} \overset{d_{\E}\to\infty}{\longrightarrow}
  & \begin{cases}0 & \left(i\neq j\right) ,\\
\sigma_{i}^{2} & \left(i=j\right) .
\end{cases} \label{eq:equi-state_of_s}
\end{eqnarray}
 Finally, substituting Eq.~(\ref{eq:hiew=00003D0}) into Eq.~(\ref{eq:change of rho(compact form)})
 and noticing Eq.~(\ref{drhoS=0}), one gets Eq.~(\ref{eq:rho_hs_commute}).

 Equation (\ref{eq:rho_hs_commute}) implies that for long times the RDM $\rho^S$
 has a diagonal form in the eigenbasis of $H_S$.
 Specifically, it can be written as
\begin{equation}\label{rho-Pa}
 \rho^S = \sum_a \chi_a P_a,
\end{equation}
 with eigenvalues $\chi_a$ ($\chi_a \ne \chi_b$ for $a\ne b$), where $P_a$ are projection operators
 composed of eigenstates of $H_S$, namely,
\begin{equation}\label{Pa}
 P_a = \sum_{|E^S_k\ra \in G_a} |E^S_k\ra \la E^S_k|,
\end{equation}
 with  ${H_S |E^S_k\ra = E^S_k|E^S_k\ra} $.
 Here, $G_a$ denotes the set of the eigenstates $|E^S_k\ra $ corresponding to the same eigenvalue
 $\chi_a$ of $\rho^S$.

 With the results obtained above, we can discuss properties of the SPB in the limit of large $d_\E$.
 Note that the Hilbert space of the system $S$ has a finite dimension $d_S$.
 If the total Hamiltonian has no symmetry that can induce degeneracy of diagonal elements of the RDM,
 usually, the diagonal elements of the RDM have finite separations.
 This implies that each set $G_a$ discussed above has one element only.
 In this case, the eigenstates of $H_S$ form a good SPB.
 Since validity of Eq.~(\ref{eq:rho_hs_commute}) is independent of the interaction strength,
 we thus get a uniform picture for the SPB in the whole coupling regime
 under the conditions specified above.

 An opposite, extreme case is also worth mentioning.
 That is, as a result of some properties of the total Hamiltonian,
 the RDM may become completely degenerate, namely, $\rho^S \to \frac{1}{d_S}{I_S}$,
 where $I_S$ is the identity operator in the Hilbert space $\HH_S$.
 This phenomenon, sometimes called depolarization of the subsystem, has been known to
 appear for a total system undergoing a uniformly random evolution \cite{Lubkin78,Lloyd88,Page93,Sen96,Sommers04,Muller12}.
 Obviously, in this case, there is in fact no basis that is ``preferred'' in view of the RDM.

 \subsection{The RDM under a generic interaction Hamiltonian}
 \label{sect-generic-HI}

 In this subsection, we show validity of Eq.~(\ref{eq:rho_hs_commute})
 under a generic interaction Hamiltonian.
 First, we discuss the following generic form of the interaction Hamiltonian,
 with a finite number of product terms, namely,
\begin{equation}\label{HI-g}
 H_I = \sum_\eta H^{IS}_\eta \otimes H^{I\E}_\eta ,
\end{equation}
 for which $\ov h_\eta^{I\E} =0$ for all values of $\eta$, where
\begin{equation} \label{hIE=0-eta}
 \ov h_\eta^{I\E} \equiv \lim_{d_{\E}\to\infty} \tr_\E H_\eta^{I\E}/d_\E .
\end{equation}
 In this case, Eq.~(\ref{eq:change of rho(compact form)}) still holds, with $W^{(2)}_{ij}$ replaced by
\begin{eqnarray}
 W^{(2)}_{ij}  = \sum_\eta \sum_{q=1}^{d_{S}}H_{\eta,qj}^{IS}H_{\eta,qi}^{I\E}-
 \sum_\eta \sum_{p=1}^{d_{S}}H_{\eta,ip}^{IS}H_{\eta,jp}^{I\E}, \label{W2}
\end{eqnarray}
 where
\begin{eqnarray} \label{HIS-eta} H_{\eta,ij}^{IS} & \eq & \la S_{i}|{H}_\eta^{IS}|S_{j}\ra,
 \\ H_{\eta,ij}^{I\E} & \eq & \la\E_{i}(t)|{H}_\eta^{I\E}|\E_{j}(t)\ra . \label{HIE-eta}
\end{eqnarray}
 We use $|\varphi_{\eta,n}\ra$ to denote eigenstates of $H^{I\E}_\eta$,
\begin{eqnarray}
\label{theta-n-eta}
 {H}^{I\E}_\eta |\varphi_{\eta,n}\ra= h_{\eta ,n}^I |\varphi_{\eta,n}\ra.
\end{eqnarray}
 Note that, since the operators $H^{I\E}_\eta$ are not necessarily commutable with each other,
 the interaction Hamiltonian $H_I$ does not necessarily have eigenvectors in the Hilbert space $\HH_\E$.
 We also assume that there exists a basis, still denoted by ${ \{|\varphi_n\ra\} } $,
 in which the coefficients $C_{in}(t)$ in the expansion in Eq.~(\ref{Ei-expan}) can be
 treated as independent random variables $X_{in}$ for each time $t>\tau_R$.

 Here, we further assume that the distributions of the real and imaginary parts of $X_{in}$
 have a {Gaussian} form, with the same variance for the same label $i$.
 Moving from the basis ${ \{|\varphi_n\ra\} } $ to a basis ${ \{|\varphi_{\eta,n}\ra\} } $,
 which is given by a unitary transformation,
 and making use of the Gaussian form of the distributions of the coefficients $C_{in}$,
 it is straightforward to verify that the coefficients of the expansion of $|\E_i(t)\ra $
 in a basis ${ \{|\varphi_{\eta,n}\ra\} } $ can also be regarded as independent random variables,
 {with properties similar to those of $C_{in}$}.
 Then, following arguments similar to those given in the previous subsection,
 it is not difficult to verify that Eq.~(\ref{eq:rho_hs_commute}) is still valid.

 Next, we discuss the case with nonzero $\ov h_\eta^{I\E}$.
 What one needs to do here is just to perform a renormalization
 to the self and the interaction Hamiltonians.
 Specifically, one may write the total Hamiltonian in Eq.~(\ref{H}) in the following form,
\begin{equation}\label{}
 H= \ww H_S + \ww H_I + H_\E ,
\end{equation}
where
\begin{eqnarray}
 \ww H_S & = & {H}_{S}+ \sum_\eta \ov h_\eta^{I\E} H^{IS}_\eta  ,
 \\  \ww H_I & = & \sum_\eta H_\eta^{IS} \otimes \ww H_\eta^{I\E} \ \ \
 \text{with} \ \ww H_\eta^{I\E} =  H_\eta^{I\E} - \ov h_\eta^{I\E} .
\end{eqnarray}
 Obviously, the values of $\tr_\E (\ww H_\eta^{I\E})/d_\E$ are zero in the limit $d_\E\to \infty$.
 Then, similar to Eq.(\ref{eq:rho_hs_commute}), one has
\begin{equation}  \label{rhos-wwHs}
 \lim_{d_{\E}\to\infty} \left[ \ww{H}_{S}, {\rho}_{S}(t)\right]=0 \quad \text{for}\ t > \tau_R
\end{equation}
 and the RDM is written as
\begin{equation}\label{rho-wwPa}
 \rho^S = \sum_a \chi_a \ww P_a,
\end{equation}
 where
\begin{eqnarray}
 \ww P_a = \sum_{|\ww E^S_k\ra \in G_a} |\ww E^S_k\ra \la \ww E^S_k|,
 \quad \text{with} \ \ww H_S |\ww E^S_k\ra = {\ww E^S_k}|\ww E^S_k\ra .
\end{eqnarray}

 In the generic case with nonzero $\ov h^{I\E}_\eta$, the discussions
 on the SPB given at the end of the
 previous subsection are still valid, with the Hamiltonian $H_S$ replaced by the renormalized one
 $\ww H_S$.
 Moreover, the value of a nonzero $\ov h_\eta^{I\E}$ gives a measure to the interaction strength.
 With increasing the interaction strength, the renormalized Hamiltonian $\ww H_S$
 changes continuously from  $H_S$ to a form {dominated by
 $\Delta H_S \equiv \sum_\eta \ov h_\eta^{I\E} H^{IS}_\eta$}.
 This implies that, if a SPB exists,
 it changes continuously from the eigenbasis of $H_S$ to the eigenbasis of $\Delta H_S$.
 In the simplest case that $\eta$ takes one value only,
 the eigenbasis of $\Delta H_S$ is just that of the interaction Hamiltonian $H_I$,

\section{Numerical results in two models}
\label{sect-numeric}

 In this section, we discuss numerical results obtained in two models,
 to illustrate the analytical results given in the previous section.
 We also discuss the complexity in identifying a good SPB,
 when dealing with an environment possessing a finite Hilbert space.

\subsection{Two models}

 We employ two models in our numerical simulation, each of which is composed of a central system $S$
 and a complex environment $\E$.
 The central system $S$ is a qubit in the first model, and is composed of
 two interacting qubits, denoted by $s$ and $A$, respectively, in the second model
 with only the qubit $A$ coupled to the environment.
 In both models, the environment is simulated by a quantum kicked rotor (QKR) in the chaotic regime.
 Recent experiments show that this type of the Hamiltonian can be realized experimentally \cite{Para13}.

 The Hamiltonian of the first model is written as
\begin{equation}\label{}
 {\mh}= {\mh}_{S}+ {\mh}_{\E}+ {\mh}_{I},
\end{equation}
where
\begin{eqnarray}
 {\mh}_{S} & = &\hbar\omega_{x}^{S} {\sigma}_{x}^{S}+\hbar\omega_{z}^{S} {\sigma}_{z}^{S},
 \\ {\mh}_{\E} & = & \frac{ {p}_{_{\theta}}^{2}}{2I_M}+k\hbar\cos( {\theta})\sum_{m}\delta(t-m\mt),
 \\ {\mh}_{I} & = & \lambda\hbar {\sigma}_{z}^{S}\otimes\cos( {\theta})\sum_{m}\delta(t-mT).
\end{eqnarray}
 Here, we write the central-system part of $\mh_I$ in a simple form, namely, $\sigma_z^S$,
 and write $\mh_S$ in a generic form.
 It proves convenient to introduce a dimensionless Hamiltonian, ${H}  = \mt^{2} \mh /{I_M}$.
 Explicitly,
\begin{equation}\label{}
    H = H_S + H_\E + H_{I},
\end{equation}
where
\begin{eqnarray} \label{HS-1}
 {H_S}  &=&\hbar_{\rm eff}\Omega_{x}^{S} {\sigma}_{x}^{S}+\hbar_{\rm eff}\Omega_{z}^{S} {\sigma}_{z}^{S},
 \\ \label{HE-1} H_\E &= & \frac{ {P}^{2}}{2}
 +k\hbar_{\rm eff}\cos( {\theta})\sum_{m}\delta(\tau-m),
 \\ \label{HI-1} H_I &= & \lambda\hbar_{\rm eff} {\sigma}_{z}^{S}\cos( {\theta})\sum_{m}\delta(\tau-m),
\end{eqnarray}
with
\begin{equation}\label{}
\hbar_{\rm eff}\eq \frac{\mt \hbar}{I_M}\ ,\ \ {P}\eq {p}_{\theta}\frac{\mt}{I_M} , \ \
\Omega\eq\omega\mt , \ \ \tau \equiv t/T.
\end{equation}

\begin{figure}
\includegraphics[scale=0.45]{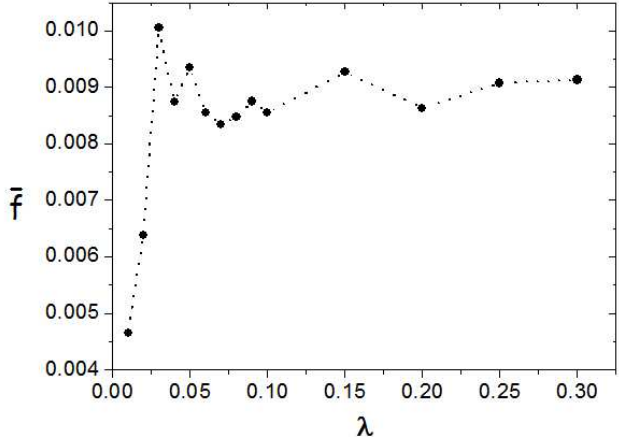}
\caption{ Variation of $\ov f$, the long-time average of the trace distance in Eq.(\ref{ft}),
 with the coupling strength in the first model.
 The average was taken over times $t\in[30\,001T,40\,000T]$.
 Parameters: $d_\E =2^{12}$, $\Omega_{x}^{S}=0.8\times10^{3}\hbar_{\rm eff}$,
$\Omega_{z}^{S}=0.8\times10^{3}\hbar_{\rm eff}$, $K=90$. }
\label{f_ave-lam-1qb}
\end{figure}

 We use the method of quantization on torus to obtain the QKR \cite{Berry80,Voros,Chirikov88,Ford91,RCB06}.
 In this scheme, $\hbar_{\rm eff}={2\pi}/{d_{\E}}$.
 The Hilbert space of the QKR is spanned by the eigenstates $|\theta_n\ra $
 of the operator $ {\theta}$, $\theta |\theta_n\ra = \theta_n |\theta_n\ra $, where
\begin{equation}\label{}
\theta_{n}=\frac{2\pi n}{d_{\E}} \quad \text{with} \ n=1,\ldots,d_{\E}.
\end{equation}
 Obviously, $|\theta_n\ra $ are eigenstates of the interaction Hamiltonian $H_I$.
 In our computation, we took the parameter ${K}\eq k\hbar_{\rm eff} =90$, for which
 the classical KR is in the chaotic regime.
 The dimensionless Schr\"{o}dinger equation is written as
\begin{equation}\label{}
i\hbar_{\rm eff}\frac{d}{d\tau}|\psi(\tau)\ra= {H}|\psi(\tau)\ra .
\end{equation}
 The Floquet operator for the time evolution within one period of time is given by
\begin{equation}
\begin{split}
 {U}_{T}   = e^{-i (\Omega_{x}^{S} {\sigma}_{x}^{S}+\Omega_{z}^{S} {\sigma}_{z}^{S})}
  e^{-i  {P}^{2}/ 2\hbar_{\rm eff}}
  e^{-i (k+\lambda {\sigma}_{z}^{S})\cos( {\theta})}.
\end{split}
\end{equation}

 \begin{figure}
 \includegraphics[scale=0.45]{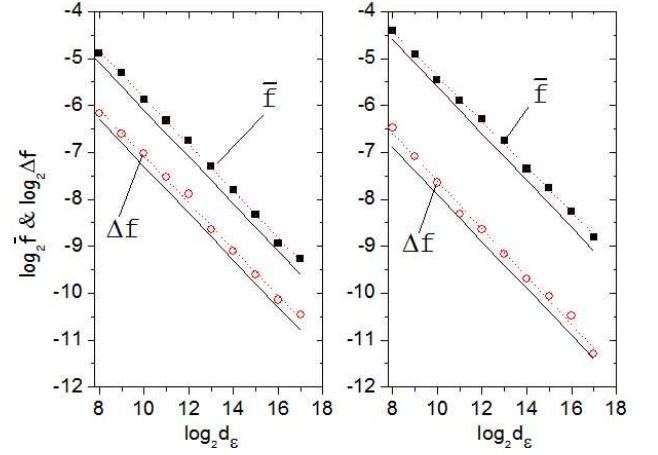}
 \caption{(Color online) Scaling behaviors of $\ov f$ (solid squares) and $\Delta f$ (empty circles)
  with the dimension $d_{\cal E}$, where $\Delta f$ is
  the standard deviation of $f(t)$ from $\ov f$. 
  Left panel: the first model with $\lambda=0.15$; right panel: the second model with $\lambda=0.1$
  and $\ve=1.0\times10^{3}$.
  The dotted straight lines give the linear fitting.
  The solid lines show straight lines with slope $-1/2$ for comparison, indicating the scaling of
  $1/\sqrt{d_{\E}}$.
}
 \label{f_nn}
 \end{figure}

 In the second model of $S+\text{QKR}=s+A+\text{QKR}$,
 the dimensionless Hamiltonian is also written as $ H = H_S + H_\E + H_{I}$,
 where $H_\E$ is the Hamiltonian of the QKR given in Eq.~(\ref{HE-1})
 and $H_S$ and $H_I$ are now written as
\begin{eqnarray}
 H_S &=& H_s +H_A + H_{sA},
 \\  H_{I} &=& { \lambda\hbar_{\rm eff} {\sigma}_{z}^{A}\cos( {\theta})\sum_{m}\delta(\tau-m) },
\end{eqnarray}
 where
\begin{eqnarray}
 & & H_s = \hbar_{\rm eff}\Omega_{x}^{s} {\sigma}_{x}^{s}+ \hbar_{\rm eff}\Omega_{z}^{s}
 {\sigma}_{z}^{s},
 \\  & & H_A = \hbar_{\rm eff}\Omega_{x}^{A} {\sigma}_{x}^{A},
 \\  & & { H_{sA} = \ve\hbar_{\rm eff} {\sigma}_{z}^{s} {\sigma}_{z}^{A} }.
\end{eqnarray}
 Note that the qubit $s$ is coupled to the qubit $A$, but is not coupled to the QKR.
 The Floquet operator in this model is written as
\begin{equation}
\begin{split}
 {U}_{T}
  = e^{-i (\Omega_{x}^{s} {\sigma}_{x}^{s}+\Omega_{z}^{s} {\sigma}_{z}^{s}+\Omega_{x}^{A} {\sigma}_{x}^{A}+\ve {\sigma}_{z}^{s} {\sigma}_{z}^{A})}
   e^{-i  {P}^{2}/ 2\hbar_{\rm eff}}
 \\ \times e^{-i (k+\lambda {\sigma}_{z}^{A})\cos( {\theta})}.
\end{split}
\end{equation}
 In the two models discussed above, the interaction Hamiltonian $H_I$ has the property
 of $\ov h^{I\E} =0 $; therefore, there is no need to consider renormalization of the self-Hamiltonian.

 \begin{figure}
 \includegraphics[scale=0.45]{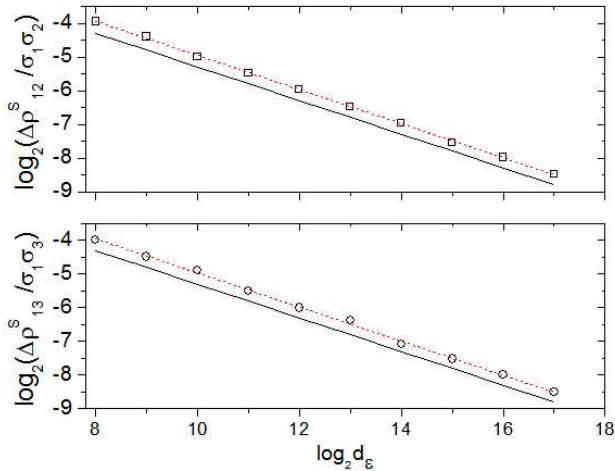}
 \caption{(Color online) Upper panel: $\Delta \rho^S_{12}/ \sigma_1 \sigma_2$
 (empty squares) versus $d_{\E}$ in the logarithm scale for the first model with $\lambda=0.15$.
 Lower panel: $\Delta \rho^S_{13}/ \sigma_1 \sigma_3$ (empty circles) in the second model
 with  $\ve=1.0\times10^{3}$ and $\lambda=0.1$.
 The solid straight lines show the analytically predicted slope $-1/2$.
}
 \label{scale}
 \end{figure}

 \subsection{Numerical results in the first model}

 Due to the chaotic motion of the QKR, it is reasonable to expect that,
 under a generic initial state of the total system, the coefficients $C_{in}(t)$ in
 Eq.~(\ref{Ei-expan}) can be regarded as random numbers effectively for long times $t>\tau_R$.
 It is still a little subtle whether the analytical result Eq.~(\ref{eq:rho_hs_commute})
 is applicable to the two models discussed above,
 because the eigenvalues $\theta_n$ approach {a} continuum in the limit of large dimension $d_\E $.
 In Appendix~\ref{app-kr}, it is shown that, in this case with a non-discrete spectrum of $H^{I\E}$,
 Eq.~(\ref{HIE-CC}) still holds; as a result, Eq.~(\ref{eq:rho_hs_commute}) is still valid.

 \begin{figure}
 \includegraphics[scale=0.43]{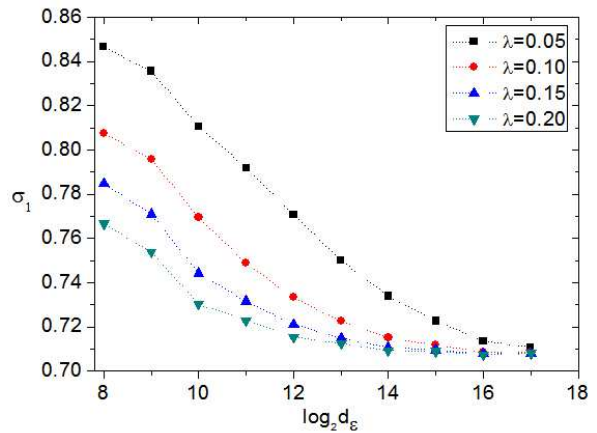}
 \caption{(Color online.)
 $\sigma_1$ versus the dimension $d_{\cal E}$ (in the logarithm scale) in the first model.
 The values of $\sigma_2$ can be determined by the relation $\sigma_1^2+\sigma_2^2=1$.}
 \label{sigma1}
 \end{figure}

 First, we discuss whether the RDM approaches an approximate stationary form for long times.
 For this purpose, we have numerically computed the following trace distance between the RDM and its
 long-time average denoted by $\ov \rho^S$, namely,
\begin{equation}\label{ft}
 f(\rho^S(t), \ov \rho^S) = \frac{1}{2} {\rm Tr } \sqrt{ (\rho^S(t) - \ov \rho^S)^\dag
 (\rho^S(t) - \ov \rho^S)}.
\end{equation}
 It was found that, beyond some initial decaying stage, $f(\rho^S(t), \ov \rho^S)$ fluctuates
 around its long-time average, denoted by $\ov f$.
 (See Fig.\ref{f_ave-lam-1qb} for some examples of the values of $\ov f$ obtained
 for $t\in[30\,001T,40\,000T]$.)
 Further, we found that $\ov f$ scales as $1/\sqrt{d_{\E}}$ with increasing $d_{\E}$, as well as $\Delta f$, the deviation of $f(\rho^S(t), \ov \rho^S)$ from $\ov f$ (Fig.~\ref{f_nn}).
 These results show that the RDM should have reached an approximate stationary form at long times.
 Below, we give more detailed discussions.

 One should note that, although the trace distance $f(\rho^S(t), \ov \rho^S)$ may approach
 zero in the limit $d_\E \to\infty$ [cf. Eq.(\ref{eq:equi-state_of_s})],
 for a finite $d_\E$, the RDM has finite fluctuations around its average;
 as a result, the trace distance $f(\rho^S(t), \ov \rho^S)$ remains finite.
 To get an estimate to $\ov f$, we note that one of the main contributions to this trace
 distance is given by the fluctuation of the off-diagonal element $\rho^S_{12}$
 from its average value which is zero.
 Making use of the expression of $\rho^S_{12}$ given in the first equality in
 Eq.~(\ref{eq:equi-state_of_s}),
 with the coefficients $C_{in}$ taken as random variables, direct derivation shows that
 the standard deviation of $\rho^S_{12}$
 is given by ${\Delta \rho^S_{12}} = \sigma_1 \sigma_2 /\sqrt{d_\E} $,
 where $\sigma_i^2$ are the variances of the random variables.

 Numerically, we checked that ${\Delta \rho^S_{12}}$ scales as $1/\sqrt{d_\E}$ (see Fig.~\ref{scale}).
 Moreover,${\Delta \rho^S_{12}}$ indeed gives the main contribution to $\ov f$.
 In fact, the values of $(\ov f - \Delta \rho^S_{12})$ were found to be smaller than $2\times 10^{-3}$
 for all the dimensions $d_\E$ studied. The dependence of $\sigma_1$ on $d_\E$ and $\lambda$ are shown in Fig.~\ref{sigma1}.

 \begin{figure}
\includegraphics[scale=0.4]{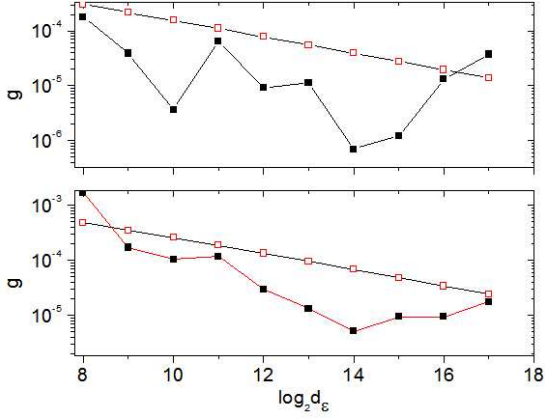}
\caption{(Color online) Values of the trace distance $g(\ov \rho^S,\ov \rho^S_e)$ (solid squares) versus $d_\E$ in the logarithm scale.
Upper panel: the first model with $\lambda=0.15$. The empty squares indicate the analytically predicted root mean square of the trace distance. Lower panel: the second model with 
$\lambda=0.1$ and $\varepsilon=10^3$. The empty squares indicate an estimated upper bound for the root mean square of the trace distance.
}
\label{d_eb_nn}
 \end{figure}

 Next, we discuss whether the averaged RDM $\ov \rho^S$ is approximately diagonal in the energy basis.
 For this purpose, we have computed the trace distance between $\ov \rho^S$ and $\ov \rho^S_e$,
 denoted by $g(\ov \rho^S,\ov \rho^S_e)$,
 where $\ov \rho^S_e$ is the diagonal part of $\ov \rho^S$ in the energy basis,
\begin{equation}\label{rho-Se}
 \ov \rho^S_e = \sum_k \la E^S_k|\ov \rho^S |E^S_k\ra |E^S_k\ra \la E^S_k|.
\end{equation}
 We found that the values of $g(\ov \rho^S,\ov \rho^S_e)$ are indeed quite small (see Fig.~\ref{d_eb_nn}),
 e.g., $g(\ov \rho^S,\ov \rho^S_e) \simeq 9.1\times10^{-6}$ for $d_\E=2^{12}$.
 This implies that $\ov \rho^S$ are approximately diagonal in the energy eigenbasis.

 To have further understanding of properties of the distance $g(\ov \rho^S,\ov \rho^S_e)$, we note that, in this model, it has the
simple expression of $g(\ov \rho^S,\ov \rho^S_e)=|\la E^S_1|\ov \rho^S |E^S_2\ra|$. 
 Let us assume that the basis $\{ |E^S_k\ra \}$ is not far from the basis $\{ |S_i\ra \}$
 and also assume that the coefficients $C_{in}(t)$ can be regarded as random numbers for each
 $t$ in the long time region, as done above when discussing $\Delta \rho^S_{12}$.
 Further, for the chaotic motion of the kicked rotator with the large parameter $K=90$,
 we assume that the correlation between the components $C_{in}(t)$
 at neighboring kicks can be neglected.
 Then, it is not difficult to see that
 $|\la E^S_1|\ov \rho^S |E^S_2\ra|$ can be approximately regarded as a random variable.
 Direct derivation shows that it has a root-mean-square $\sigma_1 \sigma_2 /\sqrt{d_{\E} N_T}$, where $N_T$ is the number of kicks within the time period for computing the average of $\rho^S$.
 Hence, usually, $0 \le g(\ov \rho^S,\ov \rho^S_e) \lesssim
 \sigma_1 \sigma_2 /\sqrt{d_{\E} N_T}$.
 For the values of $d_{\E}$ accessible in our numerical simulation,
 we have checked that the numerically computed values of $g(\ov \rho^S,\ov \rho^S_e)$ are consistent with this analytical estimate (see Fig.~\ref{d_eb_nn}).
 For example, for $d_{\E}=2^{12}$ and $N_T=10^4$,
 $(\sigma_1 \sigma_2 / \sqrt{d_{\E} N_T}) \simeq 7.8 \times 10^{-5}$,
 larger than the numerically computed value $g(\ov \rho^S,\ov \rho^S_e) \simeq 9.1\times10^{-6}$.

\begin{figure}
\includegraphics[scale=0.4]{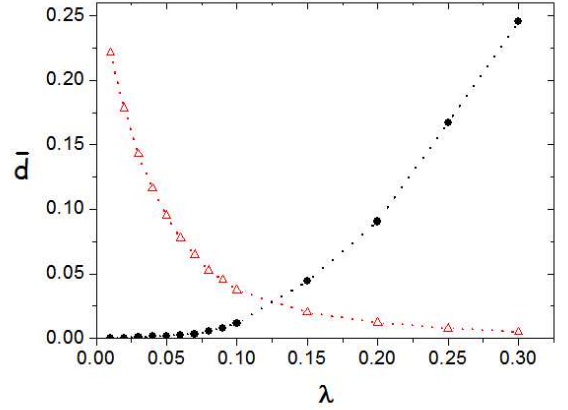}
\caption{(Color online)
 Averaged distance $\ov d$ in Eq.~(\ref{d}) in the first model (full circles),
 with $|\eta_k\ra = |E^S_k\ra $.
 The value of $\ov d$ is small for small $\lambda$, indicating that the energy eigenbasis gives a good SPB.
 The open triangles represent the trace distance between the averaged RDM $\ov \rho^S$ and
 the identity operator $I_S$ divided by $d_S$.
 }
\label{fig-d1}
\end{figure}

 Finally, we study the existence of SPB in this model.
 The finiteness of the Hilbert space of the environment makes the situation
 with the SPB more complex than that for the limit $d_\E\to \infty$ discussed in the previous section.
 In fact, as discussed above, the off-diagonal element $\rho^S_{12}$ fluctuates around
 its mean zero with a scale $\sigma_1 \sigma_2/\sqrt{d_\E}$.
 When the separation between the two diagonal elements of the RDM reduces
 to the same order of magnitude as the off-diagonal elements, namely,
 $\sigma_1 \sigma_2/\sqrt{d_\E}$, the eigenstates of the RDM may have large fluctuations,
 as  shown in Ref.\cite{WHG12}.
 In this case, the RDM has no preferred basis.
 On the other hand, in the case that the separation between the diagonal elements of the RDM is
 much larger than $\sigma_1 \sigma_2/\sqrt{d_\E}$, the energy eigenbasis gives a SPB.
 All results of our numerical simulations were found to be consistent with this understanding.

 To quantitatively characterize the extent to which eigenstates of the RDM fluctuate around their
 average, we employ a method used in Ref.~\cite{WHG12}.
 Let us use $|\rho_k(t)\ra $ with $k=1,2$ to denote eigenstates of a RDM $\rho^{S}(t)$
 with eigenvalues $\rho_k(t)$, $\rho^{S}(t) |\rho_k(t)\ra = \rho_k(t) |\rho_k(t)\ra$,
 and use $\{ |\eta_k\ra \}$ to denote an arbitrary basis.
 Since $|\la\rho_{k}(t)|\eta_{1}\ra|^{2} + |\la\rho_{k}(t)|\eta_{2}\ra|^{2}=1$,
 a good measure to the ``distance'' between the two bases is given by
\begin{equation}\label{dt}
 d(t)=1-|\la\rho_{k}(t)|\eta_{k'}\ra|^{2},
\end{equation}
 where $k$ and $k'$ are determined by the condition $|\la\rho_{k}(t)|\eta_{k'}\rangle|^{2}\ge1/2$.
 We use $\ov d$ to denote the average of this distance over a time period $[t_a,t_b]$,
\begin{equation}\label{d}
 \ov d =\frac{1}{t_{b}-t_{a}}\int_{t_{a}}^{t_{b}}dt'\ d(t').
\end{equation}
 If the value of $d$ remains small for times beyond some initial period, then, the basis
 ${ \{|\eta_{k}\ra\} } $ is {a good} SPB;
 on the other hand, large values of $d$ imply that no SPB exists.

 Our numerical computation shows that, for small values of $\lambda$,
 a parameter proportional to the interaction strength, the averaged distances $\ov d$ are small
 with $|\eta_k\ra $ taken as the energy eigenstates (full circles in Fig.~\ref{fig-d1}),
 indicating that the energy eigenbasis gives a good SPB.
 While, for large values of $\lambda$, large fluctuations of the eigenstates of the RDM
 were observed for long times, giving large values of the distance $\ov d$, hence there is no SPB.
 As discussed above, the loss of SPB is expected to be related to the emergence of
 (approximate) degeneracy in the eigenvalues of the RDM; therefore, we have also studied
 the trace distance between $\ov \rho^S$ and the identity operator $I_S$ (divided by $d_S$).
 Indeed, this trace distance was found to be large for small $\lambda$ and
 small for large $\lambda$ (open triangles in Fig.\ref{fig-d1}).

 \subsection{Numerical results in the second model}

\begin{figure}
\includegraphics[scale=0.4]{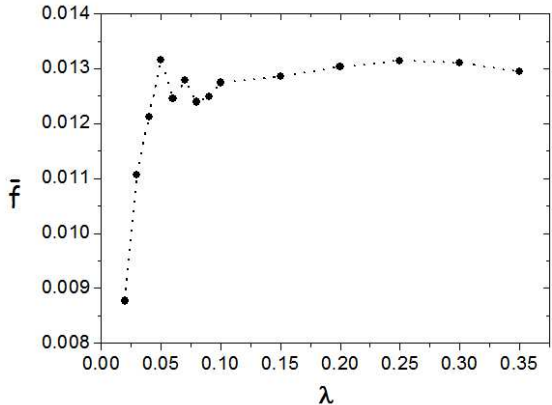}
\caption{Same as in Fig.~\ref{f_ave-lam-1qb} for the second model.
 Parameters: $\Omega_{x}^{s}=0.5\times10^{3}\hbar_{\rm eff}$,
$\Omega_{z}^{s}=1.0\times10^{3}\hbar_{\rm eff}$, $\Omega_{x}^{A}=1.5\times10^{3}\hbar_{\rm eff}$,
$\ve=1.0\times10^{3}$.
}
\label{fig-f(t)2}
\end{figure}

 In the second model of $s$+$A$+QKR, we also found numerically that the RDM of
 the central system $S=s+A$ approaches approximately stationary forms, given by $\ov\rho^S$.
 Specifically, see Fig.~\ref{fig-f(t)2} for $\ov f$, the right panel of Fig.~\ref{f_nn} for
 the scaling behavior of $\ov f$ with increasing $d_{\E}$,
 and the lower panel of Fig.~\ref{scale} for a deviation $\Delta \rho^S_{ij}$.
 Furthermore, the trace distance $g(\ov \rho^S,\ov \rho^S_e)$ is also small (see the lower panel of Fig.~\ref{d_eb_nn}),
  showing that $\ov\rho^S$
 is approximately diagonal in the energy eigenbasis.

 In this model, the trace distance $g(\ov \rho^S,\ov \rho^S_e)$ does not have an
 expression as simple as that in the first model discussed above.
 Under the same assumptions as those used in the previous subsection when discussing this trace distance,
 we can get an upper bound to its root mean square in this model.
 Specifically, making use of the inequality
 $g(\ov \rho^S-\ov \rho^S_e)\le \frac {1} {2} \sqrt{d_S}\|\ov \rho^S-\ov \rho^S_e\|_{HS}$, 
 where $\|\ov \rho^S-\ov \rho^S_e\|_{HS}$ is the Hilbert-Schmidt norm defined by
 $\|\ov \rho^S-\ov \rho^S_e\|_{HS} \eq \sqrt{\tr\{(\ov \rho^S-\ov \rho^S_e)^\dagger(\ov \rho^S-\ov \rho^S_e)\}}$,
 after some simple derivation, we find that, usually,
\begin{equation}\label{}
 0\le g(\ov \rho^S-\ov \rho^S_e) \lesssim \frac{1}{2}\sqrt{\frac{d_S}{N_T d_\E}}
 \sqrt{\sum_{i\neq j}\sigma_i^2\sigma_j^2}.
\end{equation}
 Our numerical results are consistent with this prediction
 (see the lower panel of Fig.~\ref{d_eb_nn}).

\begin{figure}
\includegraphics[scale=0.4]{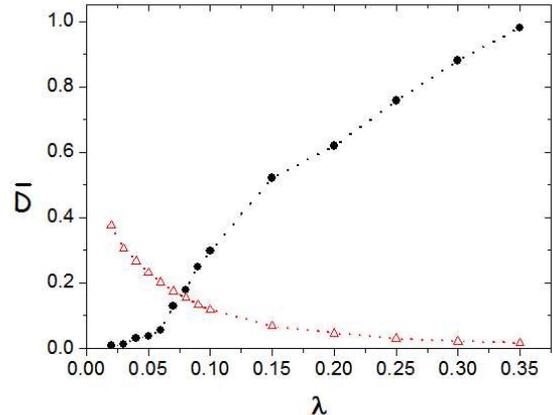}
\caption{(Color online) Similar to Fig.~\ref{fig-d1}, but for $\ov D$ (full circles),
 the long-time average of the quantity $D(t)$ in Eq.~(\ref{D}), in the second model.
 }
\label{d__dipolar-lam_2qb}
\end{figure}

 Next, we discuss the SPB. In this model with $d_S>2$,
 the quantity $d(t)$ in Eq.(\ref{dt}) is no longer a good measure to the distance between two bases.
 Therefore, we adopt a more generic measure to the distance, which is applicable for
 an arbitrary value of $d_S$, namely,
\begin{eqnarray} \label{D}
D(t)  =  \max_{k} \left\{ -\sum_{k^{\prime}}|\la\eta_{k^{\prime}}|\rho_{k}(t)\ra|^{2}\ln |\la\eta_{k^{\prime}}|\rho_{k}(t)\ra|^{2} \right\}. \ \
\end{eqnarray}
 The value of $e^{D(t)}$ gives effectively the (maximum) width of the distribution of
 the states $|\rho_k(t)\ra $ in the basis ${ \{|\eta_{k'}\ra\} } $.
 In the case that the two bases ${ \{|\rho_k(t)\ra\} }$ and ${ \{|\eta_{k'}\ra\} } $
 coincide, $D(t)=0$,
 while the maximum value of $D(t)$ is ${\ln d_S}$.
 Numerically, it was found that, for sufficiently small values of $\lambda$, the quantity $D(t)$,
 with ${ \{|\eta_k\ra\} } $ taken as the energy eigenstates ${ \{|E^S_k\ra\} } $, remains small for long times,
 indicating that the eigenbasis of $H_S$ is a good SPB (see Fig.~\ref{d__dipolar-lam_2qb}).
 While, for large values of $\lambda$, $\ov D$ is large and the RDM $\rho^S(t)$ is close to the identity
 operator $I_S$ (divided by $d_S$) for long times, showing that there is no good SPB.

\begin{figure}
\includegraphics[scale=0.42]{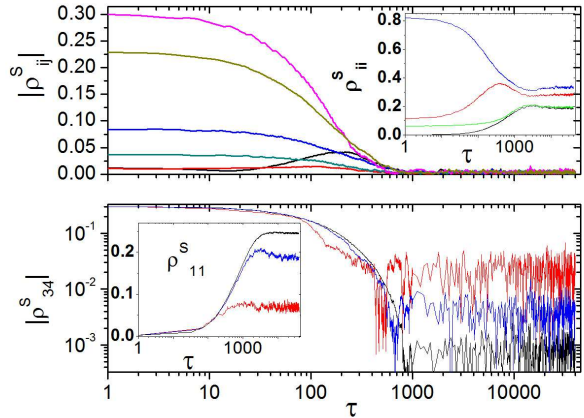}
\caption{(Color online.)
  Variation, with the rescaled time $\tau$ (in the logarithm scale),
  of elements of the RDM $\rho^S$ in the energy basis $|E^S_k\ra $
  in the second model with $\lambda=0.1$ and $\varepsilon =10^3$. Upper panel: the six off-diagonal elements
  and the four diagonal elements (inset) for $d_{\cal E}=2^{12}$.
  Two of the diagonal elements are almost degenerate beyond an initial period of time.
  Lower panel: $|\rho^{S}_{34}|$ (logarithm scale)
  for $d_{\cal E}$= $2^8$, $2^{12}$, and $2^{16}$ from top to bottom.
  Inset: $\rho^{S}_{11}$ for the same values of $d_{\E}$ (from bottom to top).
}
\label{hybrid_2qb}
\end{figure}

 In Fig.~\ref{d__dipolar-lam_2qb}, it is seen that the increasing rate of $\ov D$ for $\lambda
 $ between $0.06$ and 0.1 is obviously larger than that for $\lambda < 0.06$.
 Detailed study shows that this is related to the fact that two diagonal elements of the RDM
 become close to each other when  $\lambda $ exceeds $0.06$ (see the inset in the upper panel of Fig.~\ref{hybrid_2qb}).
 In Fig.~\ref{hybrid_2qb}, we also present some examples of the influence of the dimension $d_{\E}$ in the time evolution of the elements of the RDM. 
 It is seen that, as expected, the off-diagonal elements of the RDM decay to smaller values for
 larger dimension $d_{\E}$ and fluctuations of the elements decrease with increasing $d_{\E}$.
 Moreover, the relaxation time of the central system is not sensitive to the value of $d_\E$.

 Finally, we discuss a result given in Ref.~\cite{WHG12}. That is, taking the qubit
 $s$ as the central system and $A+$QKR as the environment, obvious deviation of
 some numerically obtained SPB from the eigenbasis of $H_s$ was found for $\lambda =0.1$.
 This does not conflict with results given in this paper, because
 for $\lambda =0.1$ the composite system $A$+QKR does
 not have a completely chaotic motion and the coefficients $C_{in}(t)$ can not be treated
 as independent random variables; as a result, Eq.~(\ref{eq:rho_hs_commute}) is not applicable.
 Indeed, the stationariness of $\rho_{s+A}=\tr_{\rm QKR} (\rho)$ implies the stationariness
 of $\rho_s = \tr_{A+{\rm QKR}} (\rho)$,
 but does not require a diagonal form of $\rho_s$ in the {eigenbasis} of $H_s$.
 We have checked numerically that the stationary RDM $\rho_{s+A}$ in this model
 indeed gives the SPB of the qubit $s$ discussed in Ref.~\cite{WHG12}.

\section{Conclusions}
\label{sect-conclusion}

 In this paper, the RDM has been studied for an open quantum system interacting with a
 quantum chaotic environment.
 It is shown that, if the RDM may become stationary beyond some time period,
 the stationary RDM has a diagonal form in the eigenbasis of a renormalized self-Hamiltonian
 of the open system, in the limit of large Hilbert space of the environment.
 Here, the renormalized self-Hamiltonian is given by the sum of the unperturbed self-Hamiltonian
 and a certain average of the interaction Hamiltonian over the degrees of freedom of the environment.
 In the case that the stationary RDM has nondegenerate eigenvalues, the eigenbasis of the
 renormalized self-Hamiltonian supplies a SPB (statistically preferred basis).
 The analytical results have been illustrated by numerical simulations performed in two models.

 The above discussed results should be useful in the study of statistical descriptions
 of open quantum systems.
 In particular, a uniform picture is available for SPB of the system,
 under the conditions specified above,
 irrespective of the strength of the system-environment interaction.
 Numerical simulations performed in the two models
 show that the situation with SPB may become more complex
 under an environment with a finite Hilbert space.
 In this case, good SPB may be destroyed, when eigenvalues of the stationary RDM become approximately
 degenerate; this may happen when the system-environment interaction becomes strong.

 Finally, we would remark that there may exist situations in which the requirement of a completely
 chaotic environment discussed above is too strong.
 In fact, what is exactly needed in the derivation of the main analytical result of this paper is that
 for each time $t$ in the long-time region 
 the coefficients $C_{in}(t)$ can be effectively regarded as independent random variables
 with certain properties.

\begin{acknowledgments}

 The authors are grateful to Jiangbin Gong for valuable discussions.
 This work was partially supported by the Natural Science Foundation of China under Grants
 No.~11275179 and No.~10975123,  the National Key Basic Research Program of China under Grant
 No.2013CB921800, and the Research Fund for the Doctoral Program of Higher Education of China.

\end{acknowledgments}

\appendix

\section{DERIVATION OF EQ.~(\ref{eq:change of rho(compact form)})}
\label{appendix-rhos}

 In this appendix, we derive Eq.~(\ref{eq:change of rho(compact form)}).
 Note that the set $\{{\bf e}_{ij}=|S_{i}\ra\la S_{j}|\}$ gives
 a basis for Hermitian operators acting on the Hilbert space of the subsystem $S$.
 In this basis, the reduced density operator can be written as
\[
{\rho}^{S}(t)=\sum_{i,j}\rho_{ij}^{S}(t){\bf e}_{ij}.
\]
 Hence $\rho_{i,j}^{S}(t)$ can be written as
\begin{equation}\label{rhosij-2}
 \rho_{i,j}^{S}(t) = \tr_{S}({\rho}^{S}(t){\bf e}_{ji}).
\end{equation}
 Making use of Eq.~(\ref{rhosij-2}) and Eq.~(\ref{drhos-dt}), one has
\begin{eqnarray}\label{rij-1}
  -i \hbar \frac{d\rho_{i,j}^{S}(t)}{dt}
  =\tr \left ( [{\rho}(t),{H}]{\bf e}_{ji}\otimes {I}_{\E} \right) ,
\end{eqnarray}
 where $I_\E$ is the identity operator in $\HH_\E$.
 The right-hand side of Eq.~(\ref{rij-1}) can be written in the following form:
\begin{equation}\label{} \nonumber
 \tr( {\rho}(t)[{H},{\bf e}_{ji}\otimes {I}_{\E}]).
\end{equation}
 Then, making use of Eqs.~(\ref{H}) and (\ref{psi-expan}), it can be shown that
\begin{eqnarray}\label{rij-2}
   -i \hbar\frac{d\rho_{i,j}^{S}(t)}{dt} = F_1 +F_2,
\end{eqnarray}
 where
\begin{eqnarray}
 F_1 = \tr \left( \sum_{p,q=1}^{n}|S_{p}\ra\la S_{q}|\otimes|\E_{p}(t)\ra\la\E_{q}(t)|
 [{H}_{S},{\bf e}_{ji}\otimes {I}_{\E}] \right),\ \
 \\ F_2 = \tr \left( \sum_{p,q=1}^{n}|S_{p}\ra\la S_{q}|\otimes|\E_{p}(t)\ra\la\E_{q}(t)|
 [ {H_{I}} ,{\bf e}_{ji}\otimes {I}_{\E}] \right). \ \
\end{eqnarray}

 Performing partial trace on the environmental part of the quantity $F_1$ and making use of
 Eq.~(\ref{rhosij-Eij}), one has
\begin{eqnarray}
  F_1 = \sum_{q=1}^{n}\la S_{q}|{H}_{S}|S_{j}\ra\rho^{S}_{iq}(t)
 - \sum_{p=1}^{n}\la S_{i}|{H}_{S}|S_{p}\ra\rho^{S}_{pj}(t) .
\end{eqnarray}
 Similarly, making use of Eq.~(\ref{HI}), one finds
\begin{eqnarray}
 F_2 =  \sum_{q=1}^{n}\la S_{q}|{H}^{IS}|S_{j}\ra\la\E_{q}(t)|{H}^{I\E}|\E_{i}(t)\ra
  \\  -\sum_{p=1}^{n}\la S_{i}|{H}^{IS}|S_{p}\ra\la\E_{j}(t)|{H}^{I\E}|\E_{p}(t)\ra  .
\end{eqnarray}
 Putting the above results together, one gets Eq.~(\ref{eq:change of rho(compact form)}).

\section{PROOF OF EQ.(\ref{HIE-CC})}
\label{app-t2}

 To prove Eq.~(\ref{HIE-CC}), we make use of the following theorem (see Ref.~\cite{Petrov}).

\noindent \textbf{Theorem 1}. Let $\left\{ Y_{n}\right\} $ be a sequence of independent,
{but not necessarily identical},
random variables with mean zero. If $a_{n}$ increases monotonically and goes to infinity
as $n \to \infty$, and
\begin{equation}\label{T1-cond}
\sum_{n=1}^{\wq}\frac{\la |Y_{n}|^{p}\ra }{a_{n}^{p}}<\wq
\end{equation}
for some $p \in [1,2]$, then
\begin{equation}\label{}
 \lim_{m \to \infty} \frac 1{a_m} \sum_{n=1}^m Y_n =0, \qquad \text{a.s.}
\end{equation}
Here, ``a.s.'' (almost sure) means that the convergence is {pointwise} and almost everywhere in the space of events (with { probability one}).

 First, we show that, in the case of $i\neq j$,
\begin{equation}
 \lim_{d_{\E}\to\infty} \frac{1}{d_{\E}} \sde h^I_n X_{jn}^{*}X_{in} = 0. \label{eq:hiew_ij}
\end{equation}
 Let us consider $Y_n=\hhwn nI {\cal R}e(X_{jn}) {\cal R}e(X_{in})$.
 Due to the independence and randomness of $X_{jn}$ and $X_{in}$,
 $Y_n$ are independent random variables with mean zero.
 Furthermore, we note that $\la|Y_n|^{2}\ra= \frac 14 |\hhwn nI|^2 \la |X_{jn}|^2
 \ra \la |X_{in}|^2 \ra  = \frac 14 |h^I_n|^2 \sigma_j^2 \sigma_i^2$.
 Hence, due to the boundedness of $H^{I\E}$, the condition (\ref{T1-cond}) is satisfied
 for $a_n =n$ and $p=2$.
 Then, from Theorem 1 we get
\begin{equation}
 \lim_{d_{\E}\to\infty} \frac{1}{d_{\E}} \sde h^I_n {\cal R}e(X_{jn}) {\cal R}e(X_{in}) = 0.
 \label{hie-p}
\end{equation}
 Taking $Y_n$ as other products of the real or imaginary parts of $X^*_{jn}$
 and $X_{in}$, results similar to Eq.~(\ref{hie-p}) can be obtained;
 therefore, Eq.~(\ref{eq:hiew_ij}) holds.

 Next, for $i=j$, we note that $Y_n = \hhwn nI(|X_{in}|^2 -\sigma_{i}^{2})$
 can also be regarded as a random variable with mean zero.
 Since $\la|C_{in}|^{4}\ra$ is finite, it is seen that
 $\la|Y_n|^{2}\ra $ is bounded; hence the condition (\ref{T1-cond}) is satisfied
 for $Y_n$ with $a_n =n$ and $p=2$.
 According to Theorem 1, this implies that
\[
 \lim_{d_\E \to \infty} \frac{1}{d_{\E}}\sde\hhwn nI(X_{in}^{*}X_{in}-\sigma_{i}^{2}) =0.
\]
Then, noticing Eq.~(\ref{hIE=0}), one has
\begin{equation}
\lim_{d_\E \to \infty} \frac{1}{d_{\E}}\sde\hhwn nI X_{in}^{*}X_{in}=0 . \label{eq:hiew_ii}
\end{equation}
 Equations (\ref{eq:hiew_ij}), (\ref{eq:hiew_ii}), and (\ref{HIE-CC}) show
 that Eq.~(\ref{HIE-CC}) indeed holds.

 Finally, we note that the above results are still valid for an unbounded sequence $\{\hhwn nI\}$,
 which has no subsequence that diverges faster than $n^{s}$ with $0<s<\frac{1}{2}$.
 In fact, in this case, there always exists a positive integer $M$,
 such that for each integer $n$ satisfying $n>M$ the following relation is satisfied:
\begin{eqnarray}
 \la|X_{in}|^{4}\ra \sum_{k=n}^{\wq} \frac{(\hhwn kI)^{2}}{k^{2}} <\wq .
\end{eqnarray}
 Then, following arguments similar to those given above,
 it is seen that  Eq.~(\ref{HIE-CC}) is still valid.

\section{EQ.~(\ref{HIE-CC}) IN THE QKR MODEL \label{app-kr}}

 In this appendix, we show that Eq.~(\ref{HIE-CC}) is still valid
 with the QKR as the environment.
 For the QKR quantized on a torus, at each kick, $h^{I}_n$ is given by $\cos {\theta_n}$, where
 $\theta_{n}={2\pi n}/{d_{\E}}$ with $n=1,\ldots, d_{\E}$.
 The quantity $(h^{I}_n X_{jn}^{*}X_{in})$, as a random variable,
 has different properties for different values of $d_\E$;
 therefore, the above-mentioned Theorem 1 is not directly applicable here.

 Let us first discuss the case of $i=j$.
 In this case, since
\begin{equation}\label{}
 \lim_{d_{\E}\to\infty}\frac{1}{d_{\E}}\sum_{n=1}^{d_{\E}}\cos(\frac{2\pi n}{d_{\E}})=0 \ ,
\end{equation}
 Eq.~(\ref{HIE-CC}) can be written as
\begin{equation}
\lim_{d_{\E}\to\infty}\frac{1}{d_{\E}}\sum_{n=1}^{d_{\E}} Z_{n}\cos(\frac{2\pi n}{d_{\E}})=0, \qquad \text{a.s.},
\label{proved by toeplitz theorem}
\end{equation}
 where $Z_n = |X_{in}|^2 -\sigma_i^{2}$.
 Obviously, due to the independency and randomness of $X_{in}$,
 the quantities $Z_n$ are independent random variables with mean zero.
 Furthermore, it is readily seen that the following average of the random variables $Z_n$, namely,
\begin{equation}\label{Sk}
 S_{k}=\frac{1}{k} \sum_{n=1}^{k} Z_{n},
\end{equation}
 are also random variables with mean zero.
 According to the strong law of large numbers,
\[
\lim_{d_{\E}\to\infty}S_{d_{\E}}=0, \quad \text{a.s.}
\]
 To continue the proof, we need to make use of the second part of the following theorem \cite{Toeplitz}.

\noindent
\textbf{Theorem 2} (Toeplitz theorem). Let $\{a_{ni}\}$ be a matrix of real numbers
and $\{x_{i}\}$ a sequence of real numbers. Suppose $x_{i}\to x$ as $i\to\wq$.

\noindent (1) In the case of $x\neq0$, if
  $\sum_{i=1}^{\wq}|a_{ni}|  \le  M<\wq $  \text{for all } $n\ge1$,
 $a_{ni}\to0\ \text{as }n\text{\ensuremath{\to\wq}}$ \text{for each} $i\ge1$,
 \text{and} $\sum_{i=1}^{\wq}a_{ni}\to1\ \text{as}\ n\to\wq$,
then
\begin{equation}\label{}
\sum_{i=1}^{\wq}a_{ni}x_{i}\to x\quad\text{as}\ n\to\wq .
\end{equation}

\noindent (2) In the case of $x=0$, if
\begin{eqnarray*}
\sum_{i=1}^{\wq}|a_{ni}|  \le  M<\wq\ \quad \text{for all }n\ge1
\\ \text{and} \  a_{ni}\to0\ \text{as }n\text{\ensuremath{\to\wq}}\ \quad \text{for each}\ i\ge1 ,
\end{eqnarray*}
then
\begin{equation}\label{}
\sum_{i=1}^{\wq}a_{ni}x_{i}\to0\quad\text{as}\ n\to\wq .
\end{equation}

 Let us write the left-hand side of Eq.~(\ref{proved by toeplitz theorem}) in terms of $S_k$, i.e.,
\begin{align*}
  \lim_{d_{\E}\to\infty}\frac{1}{d_{\E}}\sum_{k=1}^{d_{\E}}(kS_{k}-(k-1)S_{k-1})\cos(\frac{2\pi k}{d_{\E}}),
\end{align*}
 then, write it further as
\begin{equation}\label{sum-S}
 \lim_{d_{\E}\to\wq}\sum_{k=1}^{\wq}a_{d_{\E}k}S_{k},
\end{equation}
 where $S_k$ is defined by Eq.(\ref{Sk}) for $k \le d_\E$ and is equal to zero for $k > d_\E$,
  and the matrix $\{a_{d_{\E}k}\}$ is defined by
\begin{eqnarray}
\begin{aligned}a_{d_{\E}k}\eq\frac{k}{d_{\E}}(b_{d_{\E}k}-b_{d_{\E}k+1}), \end{aligned}
\label{def_a}
\\
\begin{cases}
b_{d_{\E}k}=\cos(\frac{2\pi k}{d_{\E}}), & k\le d_{\E},\\
b_{d_{\E}k}=0, & k>d_{\E}.
\end{cases}
\end{eqnarray}

 According to the second part of the Toeplitz theorem, the summation in (\ref{sum-S})
 is zero (with $S_k$ taken as $x_i$).
 This would imply Eq.~(\ref{HIE-CC}), if the following two conditions are satisfied:
\begin{eqnarray}
\sum_{k=1}^{\wq}|a_{\de k}|  \le  M<\wq\quad\text{for all }\de\ge1 , \label{cond1}
\\ a_{\de k}\to0\ \text{as }\de\to\wq\quad\text{for all}\ k\ge1 . \label{cond2}
\end{eqnarray}
 In fact, from the definition of $a_{d_{\E}k}$ in Eq.~(\ref{def_a}),
 it is easy to see that, for each given $k$, the term $a_{d_{\E}k}\to0$
 as $d_{\E}\to\wq$; hence the condition (\ref{cond2}) is fulfilled.
 To show that $\sum_{k=1}^{\wq}|a_{\de k}|$ is finite as $\de\to\wq$,
 we note that, for $k<\de$,
\begin{eqnarray}
|b_{d_{\E}k}-b_{d_{\E}k+1}|
 <  \frac{2\pi}{\de}\qquad(\de\gg1).
\label{est_diff_b}
\end{eqnarray}
 From Eq.\eqref{def_a} and \eqref{est_diff_b}, one has
\begin{eqnarray}
\sum_{k=1}^{\wq}|a_{\de k}|
 <  1+\frac{2\pi}{\de^{2}}\sum_{k=1}^{\de-1}k\qquad(\de\gg1).
\end{eqnarray}
 Then, making use of the summation formula
$\sum_{k=1}^{N}k={N(N+1)}/{2}$, we have
\begin{eqnarray}
\lim_{\de \to \wq}\sum_{k=1}^{\wq}|a_{\de k}| < 1+\pi .
\end{eqnarray}
 Therefore, the condition (\ref{cond1}) is also satisfied.
 Hence Eq.~(\ref{HIE-CC}) is valid in the case of $i=j$.

 Next, Eq.~(\ref{HIE-CC}) can be proved for $i\neq j$ following arguments similar to those
 given above for $i=j$.
 The point is to show that
\begin{equation}
\lim_{d_{\E}\to\infty}\frac{1}{d_{\E}}\sum_{n=1}^{d_{\E}}Z_{1n}Z_{2n}\cos(\frac{2\pi k}{d_{\E}})=0\
 \quad (a.s.), \label{eq:HI_ij=00003D0(kr)}
\end{equation}
where $Z_{1n}$ represents ${\rm Re}(X_{jn})$ and ${\rm Im}(X_{jn})$,
and $Z_{2n}$ represents ${\rm Re}(X_{in})$ and ${\rm Im}(X_{in})$.
  In fact, introducing random variables $Z_n$ by $Z_{n}=Z_{1n}Z_{2n}$,
   Eq.~\eqref{eq:HI_ij=00003D0(kr)} can be proved following the same way as that
    given above in the proof of Eq.~(\ref{proved by toeplitz theorem}).

\end{document}